\documentclass[prc,preprint,showpacs,preprintnumbers,amsmath,amssymb]{revtex4}

\usepackage[mathscr]{eucal}
\usepackage{graphicx}
\def\slr#1{\setbox0=\hbox{$#1$}           
   \dimen0=\wd0                                 
   \setbox1=\hbox{/} \dimen1=\wd1               
   \ifdim\dimen0>\dimen1                        
      \rlap{\hbox to \dimen0{\hfil/\hfil}}      
      #1                                        
   \else                                        
      \rlap{\hbox to \dimen1{\hfil$#1$\hfil}}   
      /                                         
   \fi}

\def\ksq{k^2}

\def\mytint#1{\!\int\!\!\frac{d^3\!{#1}}{(2\pi)^3}\,}
\def\gev#1{ GeV${}^{#1}$}
\def\be{\begin{eqnarray}}
\def\ee{\end{eqnarray}}

\renewcommand{\theequation}%
    {\arabic{section}.\arabic{equation}}
\makeatletter \@addtoreset{equation}{section} \makeatother

\begin{document}

\preprint{BCCNT: 03/101/320}

\title{Quark Model Calculations of Spectral Functions of Hadronic Current Correlation Functions at Finite Temperature}

\author{Hu Li}
\author{C. M. Shakin}
 \email[email:]{casbc@cunyvm.cuny.edu}
\author{Qing Sun}

\affiliation{%
Department of Physics and Center for Nuclear Theory\\
Brooklyn College of the City University of New York\\
Brooklyn, New York 11210
}%

\date{October, 2003}

\begin{abstract}
We calculate spectral functions associated with hadronic current
correlation functions for vector and pseudoscalar currents at
finite temperature. We make use of the Nambu--Jona--Lasinio (NJL)
model with temperature-dependent coupling constants and
temperature-dependent momentum cutoff parameters. At low energies,
good fits are obtained for the spectral functions that were
extracted from lattice data by means of the maximum entropy method
(MEM). Our model has two parameters which are used to fix the
magnitude and position of the large peak seen in the spectral
functions. With those two parameters fixed, we obtain a
satisfactory fit to the width of the peak. The model then also
reproduces the energy of a second peak seen in the spectral
functions. In the case of the pseudoscalar spectral function, the
calculated peak is about 20 percent higher than that found for the
spectral function obtained from the lattice data. However, it
appears that the second peak is a lattice artifact [\,P.
Petreczky, private communication\,] and our fit to the second peak
may not be meaningful. We conclude that the NJL model may have a
broader range of application than previously considered to be the
case, if one allows for significant temperature dependence of the
parameters of the model, as well as rather large values of the
momentum cutoff parameter. Our treatment of temperature-dependent
coupling constants and cutoff parameters is analogous to the
procedure introduced by R. Casalbuoni, R. Gatto, G. Nardulli, and
M. Ruggieri, [\,Phys. Rev. D \textbf{68}, 034024 (2003)\,], who
make use of the NJL model at finite density and find that they
need to use the density-dependent coupling constants and
density-dependent cutoff parameters to study matter at high
density.
\end{abstract}


\maketitle

\section{INTRODUCTION}
In a member of recent works [\,1-\,3] we have calculated various
hadronic correlation functions and compared our results to results
obtained in lattice simulations of QCD [\,4-\,6]. The lattice
results for the correlators, $G(\tau, T)$, may be used to obtain
the corresponding spectral functions, $\sigma(\omega, T)$, by
making use of the relation \be G(\tau, T)=\int_0^\infty d \omega
\sigma_P(\omega, T) K(\tau, \omega, T)\,,\ee where \be K(\tau,
\omega, T)=\frac{\cosh[\omega(\tau-1/2T)]}{\sinh(\omega/2T)}\,.\ee
The procedure to obtain $\sigma(\omega, T)$ from the knowledge of
$G(\tau, T)$ makes use of the maximum entropy method (MEM)
[\,7-9\,], since $G(\tau, T)$ is only known at a limited member of
points.

In our previous work we have made use of the Nambu--Jona-Lasinio
(NJL) model. The Lagrangian of the generalized NJL model we have
used in our studies is

\begin{flushleft}
\be \mathcal
L=\overline{q}(i\gamma-m^0)q+\frac{G_S}{2}\sum_{i=0}^8
[(\overline{q} \lambda^{i} q)^2+(\overline{q} i \gamma_5
\lambda^{i} q)^2]\\\nonumber
-\frac{G_V}{2}\sum_{i=0}^{8}[\overline{q} \lambda^{i}\gamma_\mu
q)^2+(\overline{q} \lambda^{i}\gamma_5\gamma_\mu
q)^2]\;\;\;\;\;\;\;\;\;\;\;\;\;\;\;\;\;\;\;\;\;\;\;\\\nonumber
+\frac{G_D}{2} \lbrace
\det[\overline{q}(1+\lambda_5)q]+\det[\overline{q}(1-\lambda_5)q]\rbrace
+\mathcal L_{Conf} \ee
\end{flushleft}

Here, $m^0$ is a current quark mass matrix, $m^0=diag(m_u^0,
m_d^0, m_s^0)$. The $\lambda_i$ are the Gell-Mann (flavor)
matrices and $\lambda^0=\sqrt{2/3}$I, with I being the unit
matrix. The fourth term is the 't Hooft interaction and $\mathcal
L_{Conf}$ represents the model of confinement used in our studies
of meson properties. For the present work we neglect the 't Hooft
interaction and $\mathcal  L_{Conf}$. Thus, there are essentially
three parameters to consider, $G_S$, $G_V$ and a Gaussian cutoff
parameter $\alpha$, which restricts the momentum integrals through
a factor exp$[-\overrightarrow{k}^2/\alpha^2]$. When we use the
NJL model to study matter at finite temperature, we introduce the
temperature-dependent parameters $G_S(T)$, $G_V(T)$ and
$\alpha(T)$. These parameters are adjusted to obtain fits to the
spectral functions $\sigma(\omega, T)$ for $T/T_c=1.5$ and
$T/T_c=3.0$, which are the values of $T/T_c$ studied in the
lattice simulations of QCD that we consider in this work [\,10\,].

Our application of a generalized NJL model for the calculation of
temperature-dependent hadronic correlation functions has been
described in detail in our earlier work [\,1-3\,]. For ease of
reference, we include the relevant material in the Appendix of
this work.

In Section II we present the values obtained in our analysis of
the pseudoscalar and vector spectral functions. (The analysis of
the axial-vector and scalar spectral functions is quite similar to
that given in the Appendix for the vector and pseudoscalar
spectral functions.) In Section III we provide some further
discussion of our results and some conclusions.

\section{Calculation of  Hadronic Current Spectral Functions}

\begin{table}
\caption{Parameters used in the calculation of the results reported in Figs. 1 and 2 are given.}
\begin{tabular}{|@{\hspace{1cm}}c@{\hspace{1cm}}|@{\hspace{1cm}}c@{\hspace{1cm}}|@{\hspace{1cm}}c@{\hspace{1cm}}|@{\hspace{1cm}}c@{\hspace{1cm}}|}
\hline
$T/T_c$ & 0 & 1.5 & 3.0\\
\hline
$G_S$(\gev{-2}) & $13.49^{\,a}$ & 4.00 & 1.50\\
$G_V$(\gev{-2}) & $11.46^{\,a}$ & 7.50 & $2.90^{\,b}$\\
$\alpha$(\gev{-2}) & $0.605^{\,a}$ & 0.850 & 1.50\\
\hline
\end{tabular}\\

\begin{flushleft}
a) Values used in Ref. [\,20\,],\\
b) For the calculation reported in Fig. 3, this value is changed to $G_V=2.30$\gev{-2}.
\end{flushleft}
\end{table}

 \begin{figure}
 \includegraphics[bb=0 0 280 235, angle=0, scale=1]{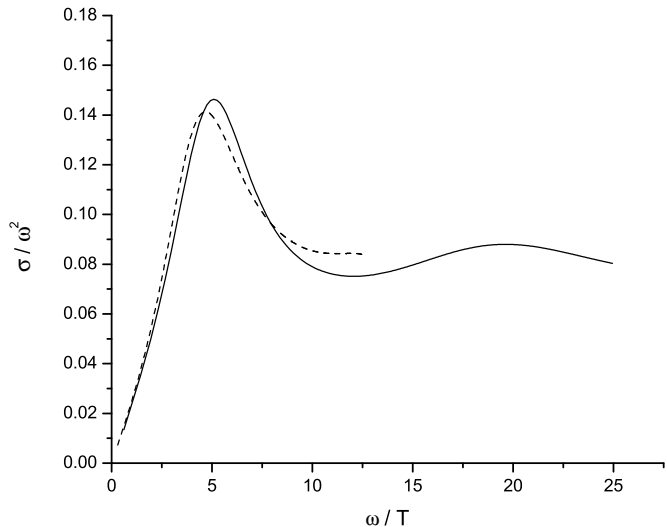}%
 \caption{Calculated values of $\sigma_P(T)/\omega^2$ are shown as a function of $\omega/T$ for the model described in the Appendix. [\,See Table 1.\,] Comparison may be made to Fig. 4 of Ref. [\,10\,] or Fig. 4 of the present work. Here the solid line corresponds to $T/T_c=1.5$, while the dashed line is calculated for $T/T_c=3.0$.}
 \end{figure}

 \begin{figure}
 \includegraphics[bb=0 0 280 235, angle=0, scale=1]{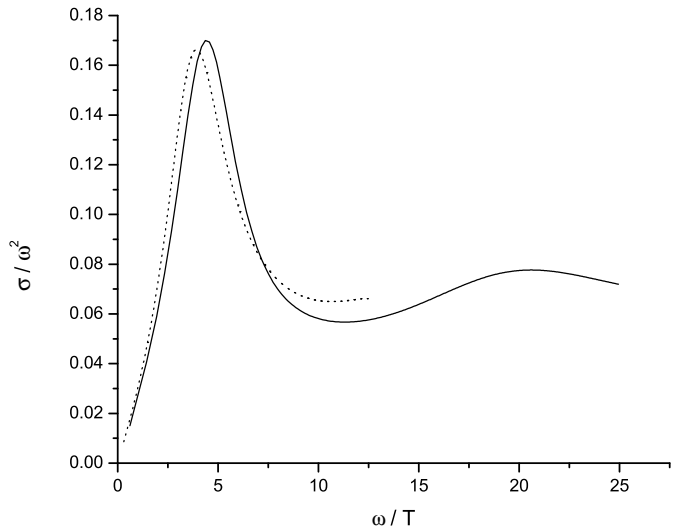}%
 \caption{Calculated values of $\sigma_V(T)/\omega^2$ are shown as a function of $\omega/T$. [\,See Table 1.\,] Here the solid line corresponds to $T/T_c=1.5$, while the dotted line is calculated for $T/T_c=3.0$. (\,See Fig. 4 of Ref. [\,10\,] or Fig. 5 of the present work\,) The theoretical value of $\sigma_V(T)$ defined in the appendix is here multiplied by $3/4$ to correspond to the normalization used in the literature [\,4-6, 10\,]. [\,See Eqs. (A30) and (A31)].}
 \end{figure}

 \begin{figure}
 \includegraphics[bb=0 0 280 235, angle=0, scale=1]{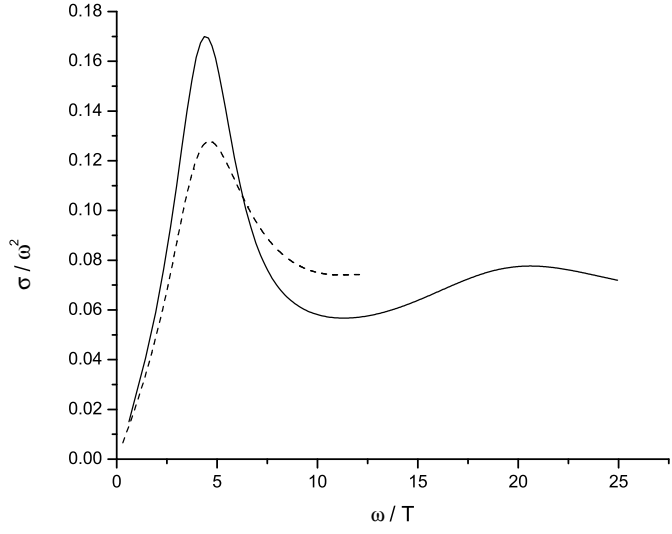}%
 \caption{Calculated values of $\sigma_V(T)/\omega^2$ are shown. The value of $G_V$ for the dashed line (\,$T/T_c=3.0$\,) is changed from the value of $2.9$ \gev{-2} used for Fig. 2 to $2.3$ \gev{-2}. See Fig. 4 of Ref. [\,10\,] or Fig. 5 of the present work. Note the large theoretical error for the vector spectral functions in Ref. [\,10\,], which makes the height of the first peak somewhat uncertain. (See the comment on the normalization of $\sigma_V(T)$ given in the caption of Fig. 2.)}
 \end{figure}

 \begin{figure}
 \includegraphics[bb=0 0 280 235, angle=0, scale=1]{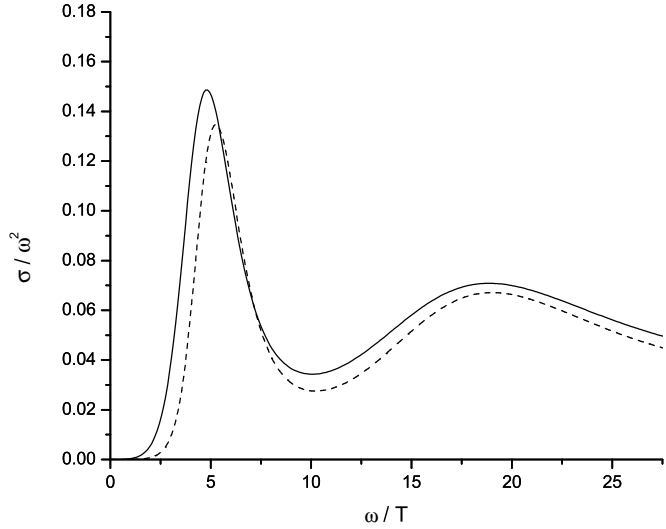}%
 \caption{The spectral function $\sigma/\omega^2$ for pseudoscalar states obtained by MEM [\,10\,] is shown. The solid line is for $T/T_c=1.5$ and the dashed line is for $T/T_c=3.0$.}
 \end{figure}

 \begin{figure}
 \includegraphics[bb=0 0 280 235, angle=0, scale=1]{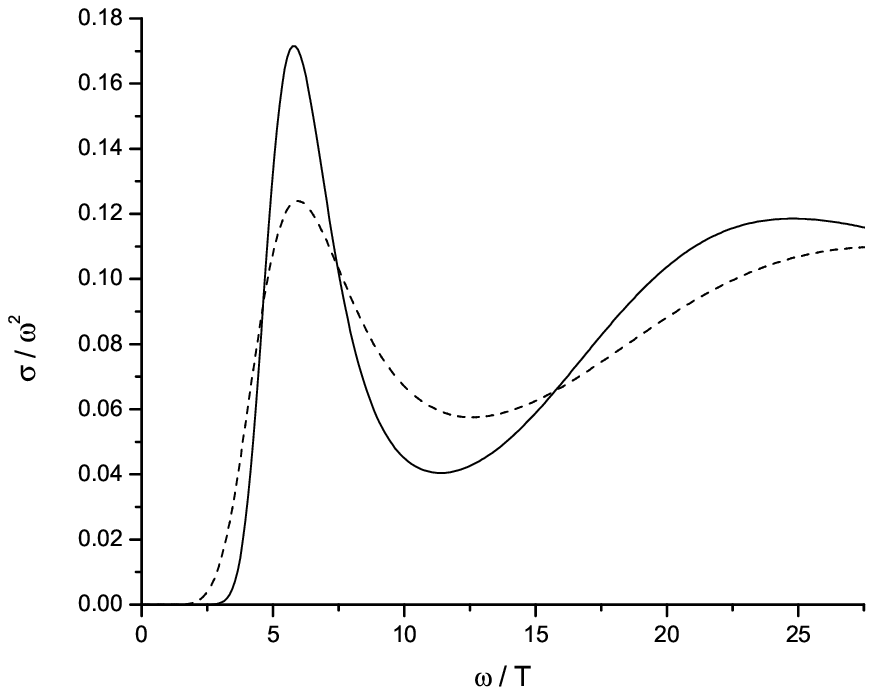}%
 \caption{The spectral function $\sigma/\omega^2$ for vector states obtained by MEM [\,10\,] is shown. See the caption of Fig. 4.}
 \end{figure}

In this section we present results obtained for
temperature-dependent hadronic current correlation functions
making use of the formalism reviewed in the Appendix. Our results
may be compared to those presented in Fig. 4 of Ref. [\,10\,].

Our calculations are made using the parameters given in Table 1.
In Fig. 1 we present values for the pseudoscalar spectral function
divided by the square of the energy, $\sigma/\omega^2$. The solid
line in Fig. 1 is obtained when $T/T_c=1.5$ and the dashed line is
for $T/T_c=3.0$. In all our calculations we use a Gaussian cutoff,
exp$[-\overrightarrow{k}^2/\alpha^2]$, when calculating the vacuum
polarization functions appearing in the denominator of the
spectral function. [\, See the Appendix.\,] The numerator of the
spectral function is calculated without a cutoff, so that our
result goes over to the perturbative result at large energies.
Since the denominator of the spectral function is calculated using
our version of NJL model, a cutoff appears naturally in the
calculation of the polarization function appearing there.

We note that, once the two parameters of the model ($G_S$ and
$\alpha$) are fixed, the width of the main peak of the
pseudoscalar spectral function seen in Fig. 1 is somewhat larger
than that seen in Fig. 4. The position of the second broad peak
seen in the spectral function obtained in the lattice study at
about 3 GeV for $T/T_c=1.5$ and at about 6 GeV for $T/T_c=3.0$ is
satisfactory. For $T/T_c=1.5$ the height of the second peak
calculated in our work is about 25 percent larger than that seen
in Fig. 4 of Ref. [\,10\,]. (\,We recall that the second peaks
obtained in the lattice simulations are thought to be
unphysical.\,)

Our results for the vector spectral function is given in Fig. 2,
where we have use the parameters of Table 1. In Fig. 4 of Ref.
[\,10\,] the height of the peak of the vector spectral function
for $T/T_c=3.0$ is significantly lower than the peak for
$T/T_c=1.5$, however, the error is quite large in this case so
that our result given in Fig. 2 is still consistent with the
lattice data. In Fig. 3, we use a smaller value of $G_V$ to lower
the peak for the $T/T_c=3.0$ curve. [\,See Table 1.\,] Both the
results given in Figs. 2 and 3 are consistent with the values
extracted from lattice data for $T/T_c=3.0$, given the large
theoretical error in this case.

As noted in the abstract, the second peak seen in the spectral
function is thought to be a lattice artifact [\,11\,]. Therefore,
the fact that we predict the energy of the second peak may not be
meaningful.

\section{conclusion}

The dynamics of the quark-gluon plasma is a topic of great
interest. It has been found that the plasma has a number of
features that cannot be described in perturbation theory. The
appearance of resonances in the temperature-dependent spectral
functions is one of these features. It may be noted in Fig. 5 of
Ref. [\,10\,] that resonances appear in the scalar, pseudoscalar,
vector, and axial-vector spectral functions at about the same
energy, which indicates that the state-dependent quark interaction
plays a less important role in separating these states in the
deconfined phase than in the confined phase. For example, it is
well known that the pion loses its character as a Goldstone boson
and becomes (approximately) degenerate with the sigma meson in the
deconfined phase.

It is of interest to note that the NJL model may be extended to
calculate resonances in the deconfined phase. In order to describe
these resonances we have made the parameters of the model, $G_S$,
$G_V$ and $\alpha$, temperature-dependent. We see that our model
provides reasonable values for the widths of the resonances and
also predicts the position of a second broad resonance seen in the
spectral functions. (However, the latter resonance is thought to
be a lattice artifact, as noted earlier.)

The temperature-dependent coupling constants and cutoff parameters
of our work are analogous to the corresponding density-dependent
parameters introduced in Refs. [\,12\,] and [\,13\,]. Further
study of models with temperature-dependent and density-dependent
parameters are of interest and a general theoretical formalism for
the introduction of such dependencies should be considered.

In future work we also hope to calculate the spatial correlators
in the deconfined phase. Some data obtained in lattice studies for
such correlation functions are given in Ref. [\,10\,].

\appendix
  \renewcommand{\theequation}{A\arabic{equation}}
  \setcounter{equation}{0}  
  \section*{APPENDIX}  

For ease of reference, we present a discussion of our calculation
of hadronic current correlators taken from Ref.\,[\,3\,]. The
procedure we adopt is based upon the real-time finite-temperature
formalism, in which the imaginary part of the polarization
function may be calculated. Then, the real part of the function is
obtained using a dispersion relation. The result we need for this
work has been already given in the work of Kobes and Semenoff
[\,14\,]. (In Ref.\,[\,14\,] the quark momentum is $k$ and the
antiquark momentum is $k-P$. We will adopt that notation in this
section for ease of reference to the results presented in
Ref.\,[\,14\,].) With reference to Eq.\,(5.4) of Ref.\,[\,14\,],
we write the imaginary part of the scalar polarization function as
\be \mbox{Im}\,J_S(\textit{P}\,{}^2,
T)=\frac12(2N_c)\beta_S\,\epsilon(\textit{P}\,{}^0)\mytint
ke^{-\vec
k\,{}^2/\alpha^2}\left(\frac{2\pi}{2E_1(k)2E_2(k)}\right)\\\nonumber
\{(1-n_1(k)-n_2(k))
\delta(\textit{P}\,{}^0-E_1(k)-E_2(k))\\\nonumber-(n_1(k)-n_2(k))
\delta(\textit{P}\,{}^0+E_1(k)-E_2(k))\\\nonumber-(n_2(k)-n_1(k))
\delta(\textit{P}\,{}^0-E_1(k)+E_2(k))\\\nonumber-(1-n_1(k)-n_2(k))
\delta(\textit{P}\,{}^0+E_1(k)+E_2(k))\}\,.\ee Here,
$E_1(k)=[\,\vec k\,{}^2+m_1^2(T)\,]^{1/2}$. Relative to Eq.\,(5.4)
of Ref.\,[\,14\,], we have changed the sign, removed a factor of
$g^2$ and have included a statistical factor of $2N_c$, where the
factor of 2 arises from the flavor trace. In addition, we have
included a Gaussian regulator, $\exp[\,-\vec k\,{}^2/\alpha^2\,]$.
The value $\alpha=0.605$ GeV was used in our applications of the
NJL model in the calculation of meson properties at $T=0$. We also
note that \be n_1(k)=\frac1{e^{\,\beta E_1(k)}+1}\,,\ee and \be
n_2(k)=\frac1{e^{\,\beta E_2(k)}+1}\,.\ee For the calculation of
the imaginary part of the polarization function, we may put
$\ksq=m_1^2(T)$ and $(k-P)^2=m_2^2(T)$, since in that calculation
the quark and antiquark are on-mass-shell. In Eq.\,(A1) the factor
$\beta_S$ arises from a trace involving Dirac matrices, such that
\be \beta_S&=&-\mbox{Tr}[\,(\slr k+m_1)(\slr k-\slr P+m_2)\,]\\
&=&2P^2-2(m_1+m_2)^2\,,\ee where $m_1$ and $m_2$ depend upon
temperature. In the frame where $\vec P=0$, and in the case
$m_1=m_2$, we have $\beta_S=2P_0^2(1-{4m^2}/{P_0^2})$. For the
scalar case, with $m_1=m_2$, we find \be \mbox{Im}\,J_S(P^2,
T)=\frac{N_cP_0^2}{4\pi}\left(1-\frac{4m^2(T)}{P_0^2}\right)^{3/2}
e^{-\vec k\,{}^2/\alpha^2}[\,1-2n_1(k)\,]\,,\ee where \be \vec
k\,{}^2=\frac{P_0^2}4-m^2(T)\,.\ee


For pseudoscalar mesons, we replace $\beta_S$ by
\be \beta_P&=&-\mbox{Tr}[\,i\gamma_5(\slr k+m_1)i\gamma_5(\slr k-\slr P+m_2)\,]\\
&=&2P^2-2(m_1-m_2)^2\,,\ee which for $m_1=m_2$ is $\beta_P=2P_0^2$
in the frame where $\vec P=0$. We find, for the $\pi$ mesons, \be
\mbox{Im}\,J_P(P^2,T)=\frac{N_cP_0^2}{4\pi}\left(1-\frac{4m^2(T)}{P_0^2}\right)^{1/2}
e^{-\vec k\,{}^2/\alpha^2}[\,1-2n_1(k)\,]\,,\ee where $ \vec
k\,{}^2={P_0^2}/4-m_u^2(T)$, as above. Thus, we see that the phase
space factor has an exponent of 1/2 corresponding to a
\textit{s}-wave amplitude. For the scalars, the exponent of the
phase-space factor is 3/2, as seen in Eq.\,(A6).

For a study of vector mesons we consider \be
\beta_{\mu\nu}^V=\mbox{Tr}[\,\gamma_\mu(\slr k+m_1)\gamma_\nu(\slr
k-\slr P+m_2)\,]\,,\ee and calculate \be
g^{\mu\nu}\beta_{\mu\nu}^V=4[\,P^2-m_1^2-m_2^2+4m_1m_2\,]\,,\ee
which, in the equal-mass case, is equal to $4P_0^2+8m^2(T)$, when
$m_1=m_2$ and $\vec P=0$. This result will be needed when we
calculate the correlator of vector currents. Note that, for the
elevated temperatures considered in this work, $m_u(T)=m_d(T)$ is
quite small, so that $4P_0^2+8m_u^2(T)$ can be approximated by
$4P_0^2$, when we consider the vector current correlation
functions. In that case, we have \be \mbox{Im}\,J_V(P^2,T) \simeq
\frac{2}{3}\mbox{Im}\,J_P(P^2,T)\,.\ee At this point it is useful
to define functions that do not contain that Gaussian regulator:
\be\mbox{Im}\,\tilde{J}_P(P^2,T)=\frac{N_cP_0^2}{4\pi}\left(1-\frac{4m^2(T)}{P_0^2}\right)^{1/2}[\,1-2n_1(k)\,]\,,\ee
and
\be\mbox{Im}\,\tilde{J}_V(P^2,T)=\frac{2}{3}\frac{N_cP_0^2}{4\pi}\left(1-\frac{4m^2(T)}{P_0^2}\right)^{1/2}[\,1-2n_1(k)\,]\,,\ee
For the functions defined in Eq.\,(A14) and (A15) we need to use a
twice-subtracted dispersion relation to obtain
$\mbox{Re}\,\tilde{J}_P(P^2,T)$, or
$\mbox{Re}\,\tilde{J}_V(P^2,T)$. For example,
\be\mbox{Re}\,\tilde{J}_P(P^2,T)=\mbox{Re}\,\tilde{J}_P(0,T)+
\frac{P^2}{P_0^2}[\,\mbox{Re}\,\tilde{J}_P(P_0^2,T)-\mbox{Re}\,\tilde{J}_P(0,T)\,]\\\nonumber
+\frac{P^2(P^2-P_0^2)}{\pi}\int_{4m^2(T)}^{\tilde{\Lambda}^{2}}
ds\frac{\mbox{Im}\,\tilde{J}_P(s,T)}{s(P^2-s)(P_0^2-s)}\,,\ee
where $\tilde{\Lambda}^{2}$ can be quite large, since the integral
over the imaginary part of the polarization function is now
convergent. We may introduce $\tilde{J}_P(P^2,T)$ and
$\tilde{J}_V(P^2,T)$ as complex functions, since we now have both
the real and imaginary parts of these functions. We note that the
construction of either $\mbox{Re}\,J_P(P^2,T)$, or
$\mbox{Re}\,J_V(P^2,T)$, by means of a dispersion relation does
not require a subtraction. We use these functions to define the
complex functions $J_P(P^2,T)$ and $J_V(P^2,T)$.

In order to make use of Eq.\,(A16), we need to specify
$\tilde{J}_P(0)$ and $\tilde{J}_P(P_0^2)$. We found it useful to
take $P_0^2=-1.0$ \gev2 and to put $\tilde{J}_P(0)=J_P(0)$ and
$\tilde{J}_P(P_0^2)=J_P(P_0^2)$. The quantities $\tilde{J}_V(0)$
and $\tilde{J}_V(P_0^2)$ are determined in an analogous function.
This procedure in which we fix the behavior of a function such as
$\mbox{Re}\tilde{J}_V(P^2)$ or $\mbox{Re}\tilde{J}_V(P^2)$ is
quite analogous to the procedure used in Ref.\,[\,15\,]. In that
work we made use of dispersion relations to construct a continuous
vector-isovector current correlation function which had the
correct perturbative behavior for large $P^2\rightarrow-\infty$
and also described the low-energy resonance present in the
correlator due to the excitation of the $\rho$ meson. In
Ref.\,[\,15\,] the NJL model was shown to provide a quite
satisfactory description of the low-energy resonant behavior of
the vector-isovector correlation function.

We now consider the calculation of temperature-dependent hadronic
current correlation functions. The general form of the correlator
is a transform of a time-ordered product of currents, \be iC(P^2,
T)=\int d^4xe^{iP\cdot x}<\!\!<T(j(x)j(0))>\!\!>\,,\ee where the
double bracket is a reminder that we are considering the finite
temperature case.

For the study of pseudoscalar states, we may consider currents of
the form $j_{P,i}(x)=\tilde{q}(x)i\gamma_5\lambda^iq(x)$, where,
in the case of the $\pi$ mesons, $i=1,2$ and $3$. For the study of
scalar-isoscalar mesons, we introduce
$j_{S,i}(x)=\tilde{q}(x)\lambda^i q(x)$, where $i=0$ for the
flavor-singlet current and $i=8$ for the flavor-octet current
[\,16\,].

In the case of the pseudoscalar-isovector mesons, the correlator
may be expressed in terms of the basic vacuum polarization
function of the NJL model, $J_P(P^2, T)$ [\,17-19\,]. Thus, \be
C_P(P^2, T)=J_P(P^2, T)\frac{1}{1-G_{P}(T)J_P(P^2, T)}\,,\ee where
$G_P(T)$ is the coupling constant appropriate for our study of
$\pi$ mesons. We have found $G_P(T)=13.49$\gev{-2} by fitting the
pion mass in a calculation made at $T=0$, with $m_u = m_d =0.364$
GeV. The result given in Eq.\,(A18) is only expected to be useful
for small $P^2$, since the Gaussian regulator strongly modifies
the large $P^2$ behavior. Therefore, we suggest that the following
form is useful, if we are to consider the larger values of $P^2$.
\be \frac{C_{P}(P^2, T)}{P^2}=\left[\frac{\tilde{J}_P(P^2,
T)}{P^2}\right] \frac{1}{1-G_P(T)J_P(P^2, T)}\,.\ee (As usual, we
put $\vec{P}=0$.) This form has two important features. At large
$P_0^2$, ${\mbox{Im}\,C_{P}(P_0, T)}/{P_0^2}$ is a constant, since
${\mbox{Im}\,\tilde{J}_{P}(P_0^2, T)}$ is proportional to $P_0^2$.
Further, the denominator of Eq.\,(A19) goes to 1 for large
$P_0^2$. On the other hand, at small $P_0^2$, the denominator is
capable of describing resonant enhancement of the correlation
function. As we have seen, the results obtained when Eq.\,(A19) is
used appear quite satisfactory. (\,We may again refer to
Ref.\,[\,15\,], in which a similar approximation is described.)

For a study of the vector-isovector correlators, we introduce conserved vector currents $j_{\mu,
i}(x)=\tilde{q}(x)\gamma_{\mu}\lambda_i q(x)$ with i=1, 2 and 3. In this case we define \be
J_V^{\mu\nu}(P^2, T)=\left(g\,{}^{\mu\nu}-\frac{P\,{}^\mu P\,{}^\nu}{P^2}\right)J_V(P^2, T)\ee and
\be C_V^{\mu\nu}(P^2, T)=\left(g\,{}^{\mu\nu}-\frac{P\,{}^\mu P\,{}^\nu}{P^2}\right)C_V(P^2,
T)\,,\ee taking into account the fact that the current $j_{\mu,\,i}(x)$ is conserved. We may then
use the fact that \be J_V(P^2,T) = \frac13g_{\mu\nu}J_V^{\mu\nu}(P^2,T)\ee and
\be\mbox{Im}\,J_V(P^2,T)&=& \frac23\left[\frac{P_0^2+2m_u^2(T)}{4\pi}\right]
\left(1-\frac{4m_u^2(T)}{P_0^2}\right)^{1/2}e^{-\vec
k\,{}^2/\alpha^2}[\,1-2n_1(k)\,]\\
&\simeq& \frac{2}{3}\mbox{Im}J_P(P^2,T)\,.\ee (See Eq.\,(A7) for
the specification of $k=|\vec k|$.) We then have \be
C_V(P^2,T)=\tilde{J}_V(P^2,T)\frac1{1-G_V(T)J_V(P^2,T)}\,,\ee
where we have introduced \be\mbox{Im}\tilde{J}_V(P^2,T)&=&
\frac23\left[\frac{P_0^2+2m_u^2(T)}{4\pi}\right]
\left(1-\frac{4m_u^2(T)}{P_0^2}\right)^{1/2}[\,1-2n_1(k)\,]\\
&\simeq& \frac{2}{3}\mbox{Im}\tilde{J}_P(P^2,T)\,. \ee
In the literature, $\omega$ is used instead
of $P_0$ [\,4-6\,]. We may define the spectral functions \be\sigma_V(\omega,
T)=\frac{1}{\pi}\,\mbox{Im}\,C_V(\omega, T)\,,\ee and \be\sigma_P(\omega,
T)=\frac{1}{\pi}\,\mbox{Im}\,C_P(\omega, T)\,,\ee

Since different conventions are used in the literature [\,4-6\,],
we may use the notation $\overline{\sigma}_P(\omega, T)$ and
$\overline{\sigma}_V(\omega, T)$ for the spectral functions given
there. We have the following relations: \be
\overline{\sigma}_P(\omega, T)=\sigma_P(\omega, T)\,,\ee and
\be\frac{\overline{\sigma}_V(\omega,
T)}{2}=\frac{3}{4}\sigma_V(\omega, T)\,,\ee where the factor 3/4
arises because, in Refs. [\,4-6\,], there is a division by 4,
while we have divided by 3, as in Eq.\,(A22).





\end{document}